What Users See – Structures in Search Engine Results Pages

*Nadine Höchstötter*

Huberverlag für Neue Medien, Department for Research & Development, Lorenzstraße 29, D – 76135 Karlsruhe, Germany, nsh@topicflux.de

*Dirk Lewandowski*

Hamburg University of Applied Sciences, Fakultät DMI, Department Information, Berliner Tor 5, D – 20099 Hamburg, Germany, dirk.lewandowski@haw-hamburg.de

**Abstract**

This paper investigates the composition of search engine results pages. We define what elements the most popular web search engines use on their results pages (e.g., organic results, advertisements, shortcuts) and to which degree they are used for popular vs. rare queries. Therefore, we send 500 queries of both types to the major search engines Google, Yahoo, Live.com and Ask. We count how often the different elements are used by the individual engines. In total, our study is based on 42,758 elements. Findings include that search engines use quite different approaches to results pages composition and therefore, the user gets to see quite different results sets depending on the search engine and search query used. Organic results still play the major role in the results pages, but different shortcuts are of some importance, too. Regarding the frequency of certain host within the results sets, we find that all search engines show Wikipedia results quite often, while other hosts shown depend on the search engine used. Both Google and Yahoo prefer results from their own offerings (such as YouTube or Yahoo Answers). Since we used the .com interfaces of the search engines, results may not be valid for other country-specific interfaces.

## Introduction

As it is widely known, the presentation of results on the search engine results pages (SERPs) heavily influences users' selection of certain results. Not only are the organic results on the first few positions preferred by the users, but also additional elements (such as advertisements) are placed around the core results list to profit from the typical user's selection behaviour. Furthermore, results that are "below the fold" (i.e., they can only be seen when a user scrolls down the results list) are seldom clicked on. This shows that it becomes increasingly important to produce results sets within the visible area ("above the fold") that are sufficient to the users' needs.

Regarding the fact that the main source of revenue of search engines are paid listings, it is interesting to evaluate their occurrence in result screens. Another fact is that search engine results screens become more sophisticated. They now also show images, news results and video results, amongst others. Additionally, search engines optimizers try to get certain pages into the first result screen. While it may have been sufficient to have a client's web page shown in the top results of the search engines, it is now essential to have them shown on one of the first few positions.

The question is what users do really see, in consideration of the mixture of results in search engines and the limited screen space available for presentation. Are there really ten or more places for results from the web index? Or are there only a few places left which are really algorithmically filled with web links? In respect to search engine optimization it is interesting if there is as much space as they think. We assume that the first listings are already taken by links from additional indexes such as the image or news index and are therefore not available for competition.

When looking at the newest data on the U.S. search market from ComScore [3], not surprisingly, Google has the largest market share, with 61.5 percent. Yahoo has 20.9 percent; Microsoft/Live.com, 9.2 percent; and Ask.com, 4.3 percent. Data for other countries show similar or even more concentrated distributions (for an overview on data for different countries, see [1]).

An interesting question is whether these market figures come from the superiority of one search engine or if there are other reasons. There are several levels in discussing this issue. The focus of our research is on quality factors for web search engines. We proposed a quality framework giving guidance on which factors should be taken into account when evaluating search engines [23]. An important element of the framework is users' satisfaction with search engines. Overall satisfaction could to a large degree result from the satisfaction with the "composition" of results pages, i.e. all the elements presented on the first results page, or narrower, on the first results screen. We refer to a results page when all results delivered by an individual search engine on its first page are considered, and to a results screen when only the results readily to been seen by a user are taken into account.

The importance of the first results screen has two reasons. Firstly, users seldom look beyond the first few results, and secondly, they only seldom use collections beyond the standard web collection (such as news, video or blog search).

Both reasons lead search engines to compose first results pages that become more and more aggregated and stuffed with different results from other collections. The aim of this paper is to shed light on the composition of search engine results pages (SERPs). We conduct an empirical investigation counting the number of appearances of different elements on the SERPs and draw conclusions what a user gets presented when using one of the major search engines.

The rest of this paper is organised as follows: First, we give an overview of the composition of SERPs and newer developments in results presentations with major Web search engines. We will explain all special results and give a schematic overview of every SERP, since every search engine presents its results page differently. The sponsored links are usually located in different areas. Then, we show how the results presentation corresponds with the typical behaviour of search engine users and how users search the Web and select their results. After that, we state the objectives of our empirical study and pose our research questions and hypotheses. In the methods section, we detail the design of our investigation. Then, we present the results and, in a separate section, discuss them in detail. In a concluding section, we emphasise the implications of our study and give suggestions for further research.

## Search engine results pages

In this section, we provide a short overview of results presentation in general Web search engines. (For a detailed discussion, see [24]).

In a SERP, different areas can be defined. The two main areas are the visible area and the scrolling area. In between, there is an imaginary fold. The visible area is what users can see immediately when opening a Web page. All information on the right and below, which is not visible, is the scrolling area.

Typically, SERPs consist of organic results (i.e., results from the general Web crawl of the search engine) and advertisements (labelled "sponsored links" or similarly; see Figure 1). This is what the user gets to see on the results page. However, as can be seen in Figure 1, additional results are shown above the organic results; in this case, there is a shortcut to look for local results. Often, users do not immediately perceive the source of these results, especially when they are not shown above the regular results, but instead mixed into the list of Web results.

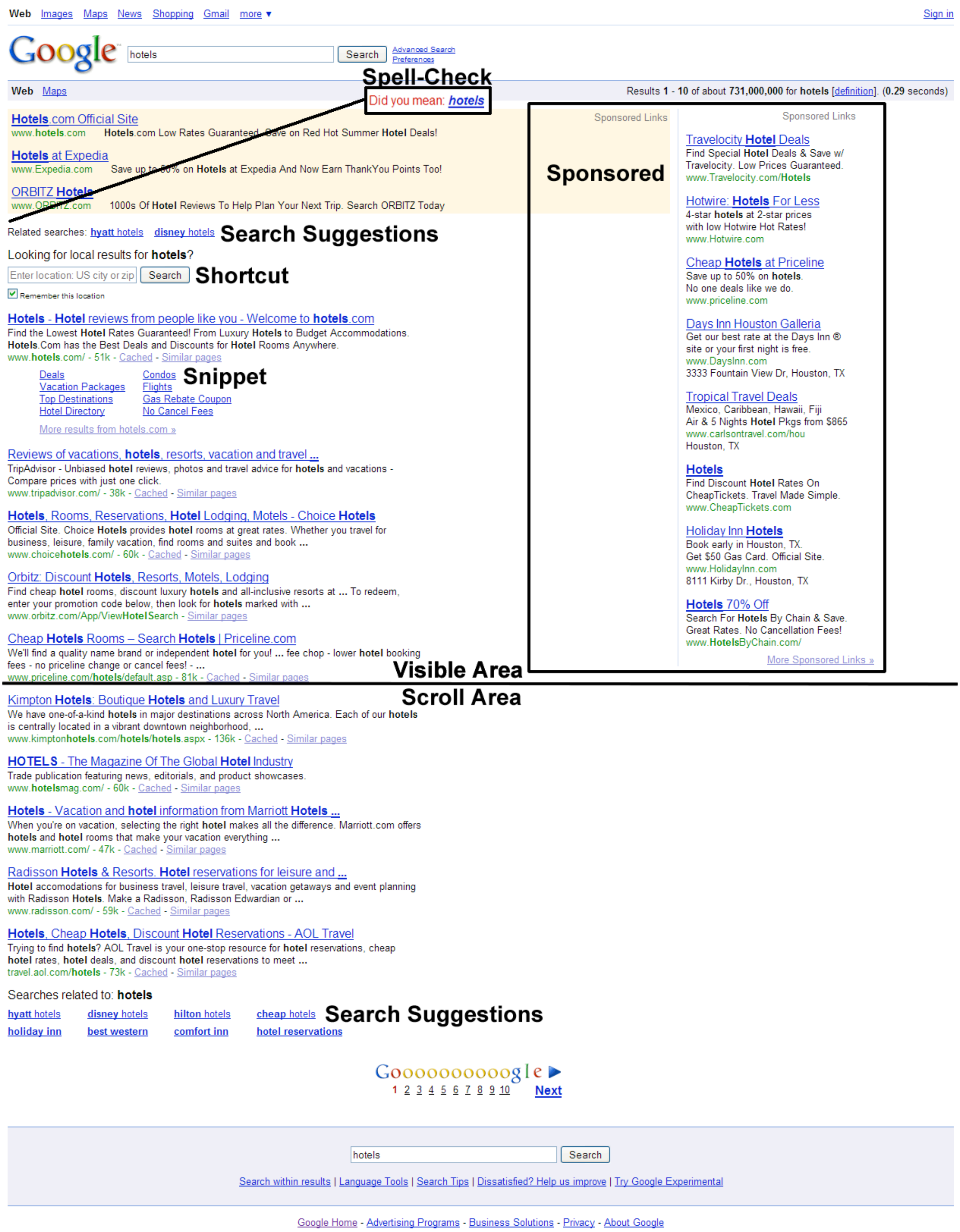

Figure 1: Example of a Google results screen, 1280x1024.

From the figure, we can also observe that the space on the results page is limited. In this case, we can see four organic listings and three advertisements on the top and eight on the right. There is also a specially presented organic result, a so-called snippet. As already mentioned, the number of results shown on the first screen depends on the screen and browser window size, respectively.

While Figure 1 shows only some elements of the results presentation in Google, we find many more depending on the query entered. As already mentioned, the central element on the results screen is the list of organic results. While search engines provide different collections (such as news, image, video, and scholarly document searches), which come from specialised Web crawls, the core of search engines remains the Web index. Different collections play more and more an important role for users. A few years ago, web pages built the biggest part of any web index. Nowadays, other media is important, too, and its portion of the web index increases rapidly. For this reason it is important to identify segments in the web index with same kind of media source, such as movies or pictures. This is helpful regarding some search queries which will likely bring up results from several indexes. Without special segments in the index, probably only web results would be ranked within the top ten. With this segmentation, search engines help users by showing automatically results from different indexes, simultaneously. Whenever a user is interested in images, only, he can also directly choose the image search interface. Web results are usually bordered with contextual ads, which are the dominant source of revenue for all major search engines, e.g., the revenues of Google in the investor relations of Q1 2008 [5], which show that the percentage of total revenue of advertising revenue is 99 percent. In the following sections, we will explain different results and their presentation on the page. We will illustrate this with screenshots and examples.

Sometimes, results are presented from another collection of the respective search engine, such as book, image, news, and blog searches. Such search results from special collections shown above the organic results are called "shortcuts" (sometimes also labelled "one-box results" or something similar). Shortcuts can also come from specialised databases, such as those containing information about patents, flights, and parcel tracking. Shortcuts are triggered by the query itself or by a special search term in the search query.

There are even more possibilities for a particular result presentation. Ask.com presents a prominent result that is a primary search result containing additional information. Primary search results are mostly triggered by the names of famous people and well-known products. They bring up an extended result with a longer description text, a picture, and some other suggestion, such as related search terms or continuative links. A primary search result from Ask.com will even be listed above the first sponsored result (Figure 2).

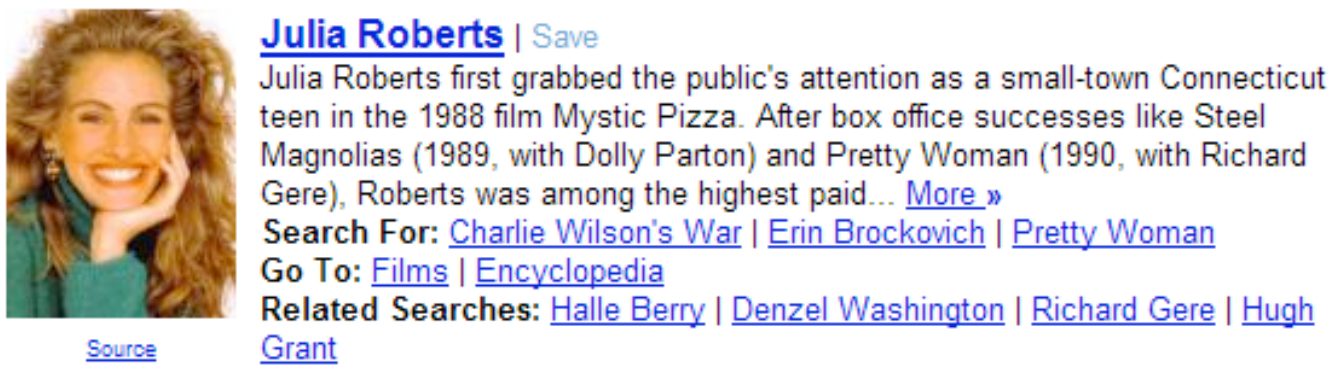

Figure 2: Example of a primary search result from Ask.com.

Users will often see results that come with a set of links located more deeply within this site. This is called a snippet. A snippet is an extended result description that shows not only the usual elements (title, short description, URL, and file size – see [22] for details) but also additional navigational links from the homepage of the Web site. Figure 3 gives an example for this from Google (note the additional search box within the snippet to search the search engine's index of that Web site directly).

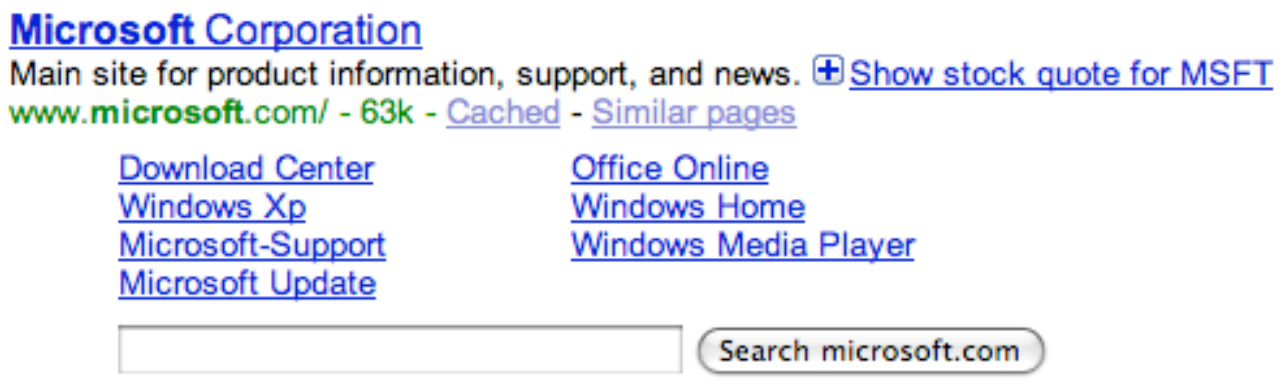

Figure 3: Snippet with additional links to the results description.

Yahoo and Google also have something smaller than a snippet but still visible in the results screen. We call this a prefetch. It gives some more results within a Web site, but those links are only listed on one line, not in two columns and several rows, as with a snippet. In Google, there is nothing special with the result, except the label "prefetch" in the HTML source code and sometimes the note "more results from..." (Figure 4).

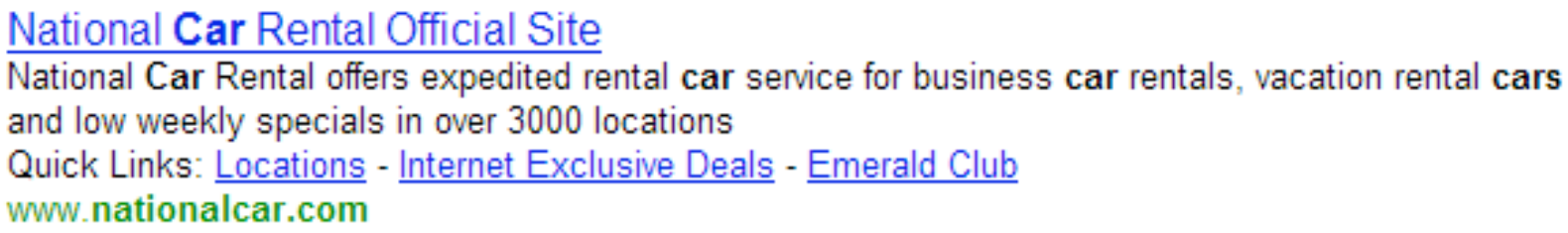

Figure 4: Prefetch result from Yahoo.

When more than one result from a Web host or site is found, search engines usually group these and only show two top results within the list. When users want to see more from the same host or site, they have to follow a link with the anchor text "More results from..." that is shown below the second result. Those second results are indented; we call such a second result a "child" of the first result.

In Table 1, we present a schematic representation of the results screens of the top four search engines. We also briefly explain the main areas of those results pages. These pages are only examples for the U.S. versions of the search engines under consideration. Results pages for different countries will differ, and the variety of search results may differ as well. The positioning and amount of sponsored links is not the same for all countries. Some search engines do not serve all countries with snippets. We concentrated our investigation on the U.S. market. In this framework, we only give examples for the US search interfaces.

Table 1: Schematic Representation of the Top Four Search Engine Results Screens

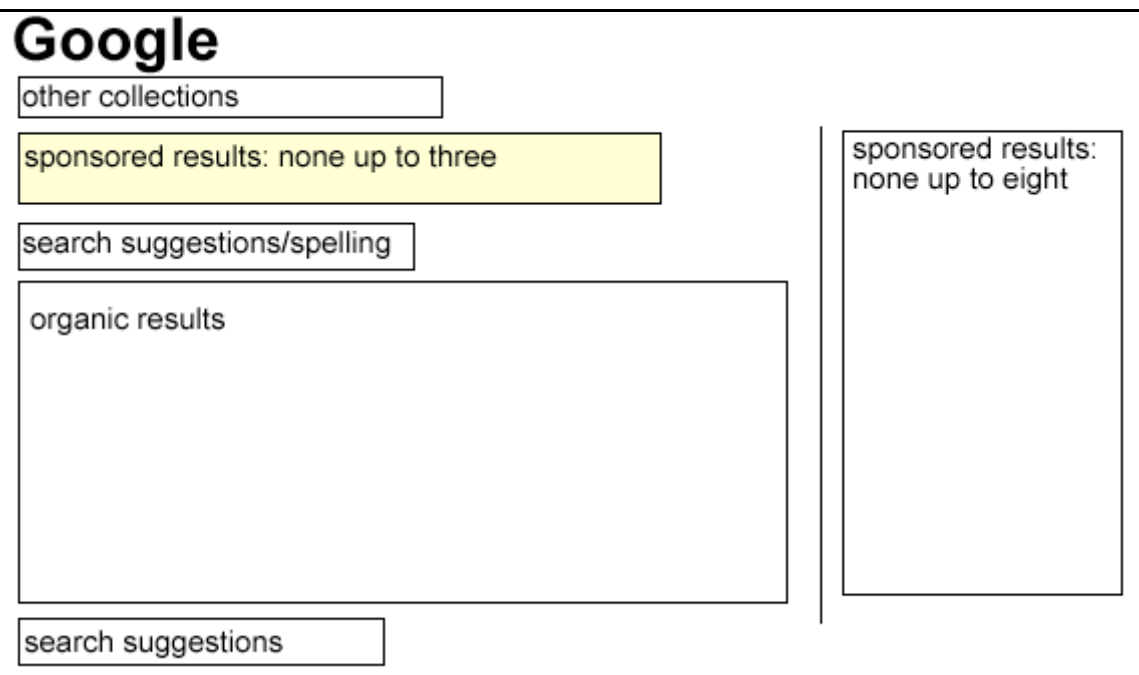

Google first gives a few links to other collections, followed by sponsored results (in a slightly coloured box), and search suggestions. Whenever there is a possible misspelling in the search query, a spelling correction will be given instead of search suggestions. Below, there are organic results. Search suggestions will again follow organic results. On the right-hand side, separated by a vertical line, are sponsored results. There is a maximum of eleven sponsored results per page.

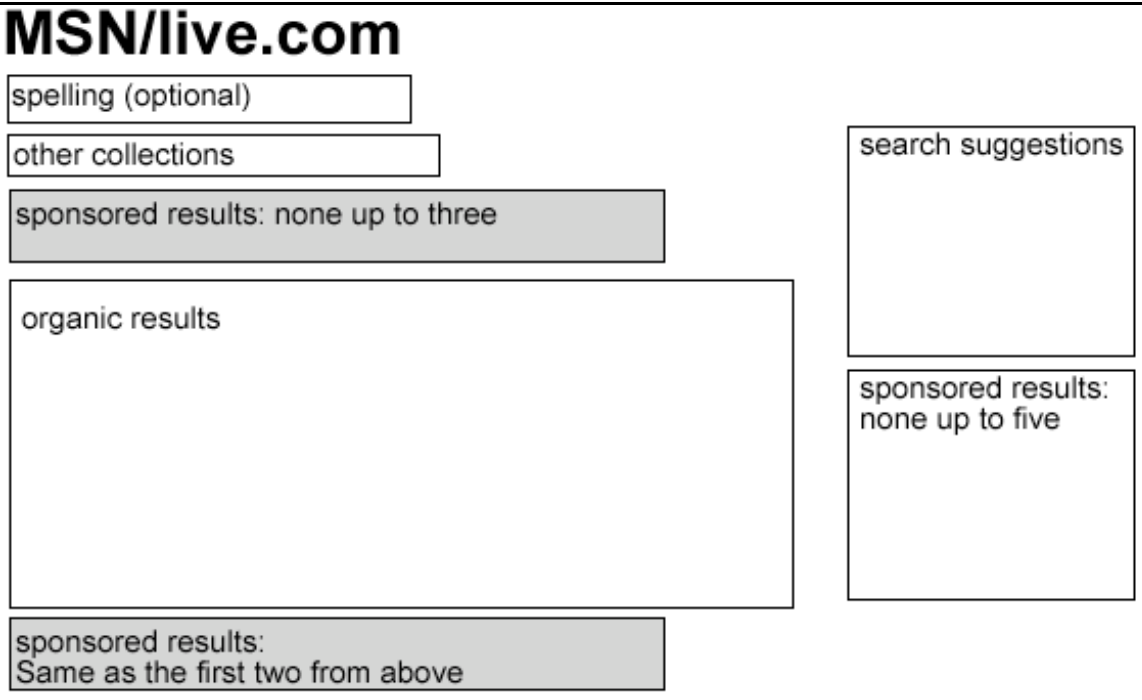

The MSN search engine has fewer fields. There are optional spelling suggestions, but the first results are normally sponsored results (in a slightly coloured box). The organic results follow. At the end, users will find the first two sponsored results from the top, again. But those are for the most common screen resolutions already in the scrolling area. On the right-hand side, there are search suggestions followed by up to five sponsored results. MSN has a maximum of ten sponsored results on the first page.

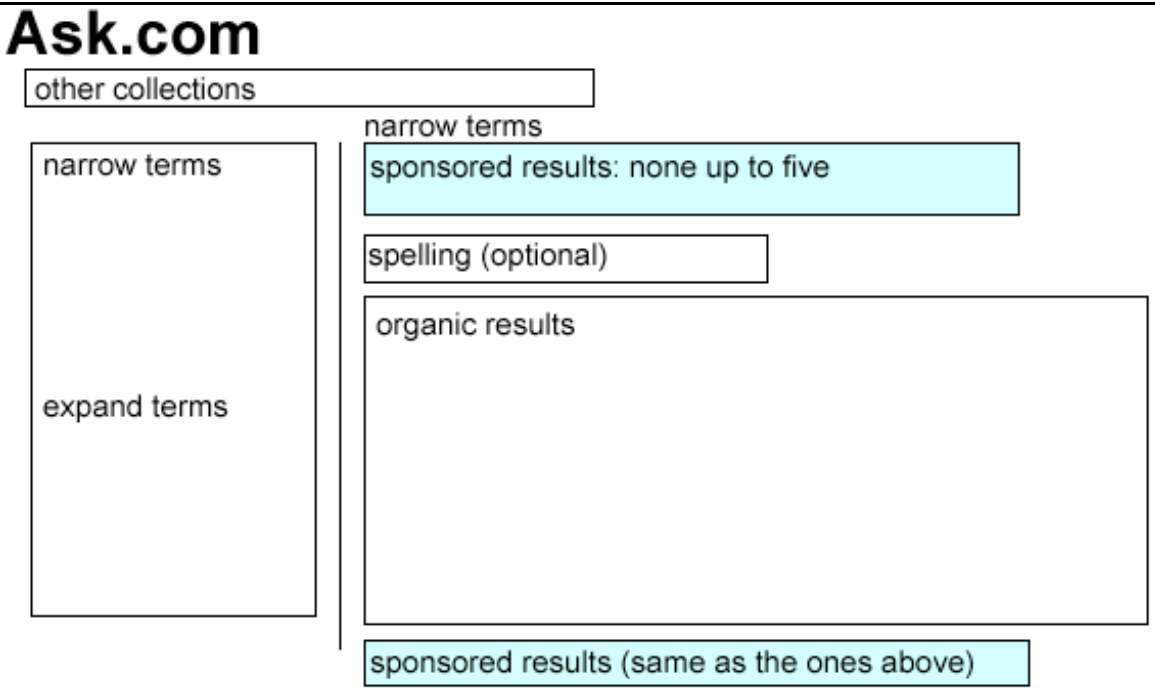

Ask.com has two different results screens. Normally, users will see links to other collections on top. On the left hand side are search suggestions to narrow or expand the query. On the other side of the vertical line are the common results, such as sponsored links. Ask.com has a maximum of ten sponsored links on the first page, but only five unique sponsored results, since the ones below are normally the same as above. Sometimes, there is a slight difference.

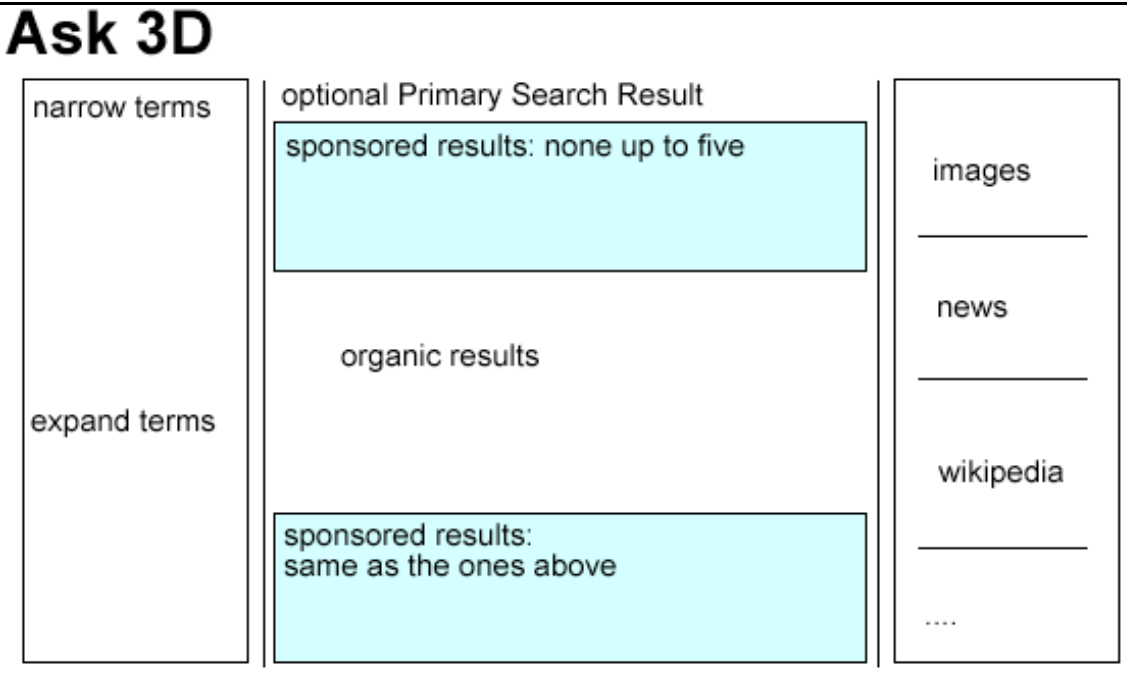

Some search queries, such as queries with a commercial touch or names of celebrities, trigger an Ask3D screen. This results screen has three columns. On the left side are search suggestions. In the middle are sponsored and organic results, and, on the right side, results from other collections. Whenever there is a primary search result, it will be above the first sponsored link.

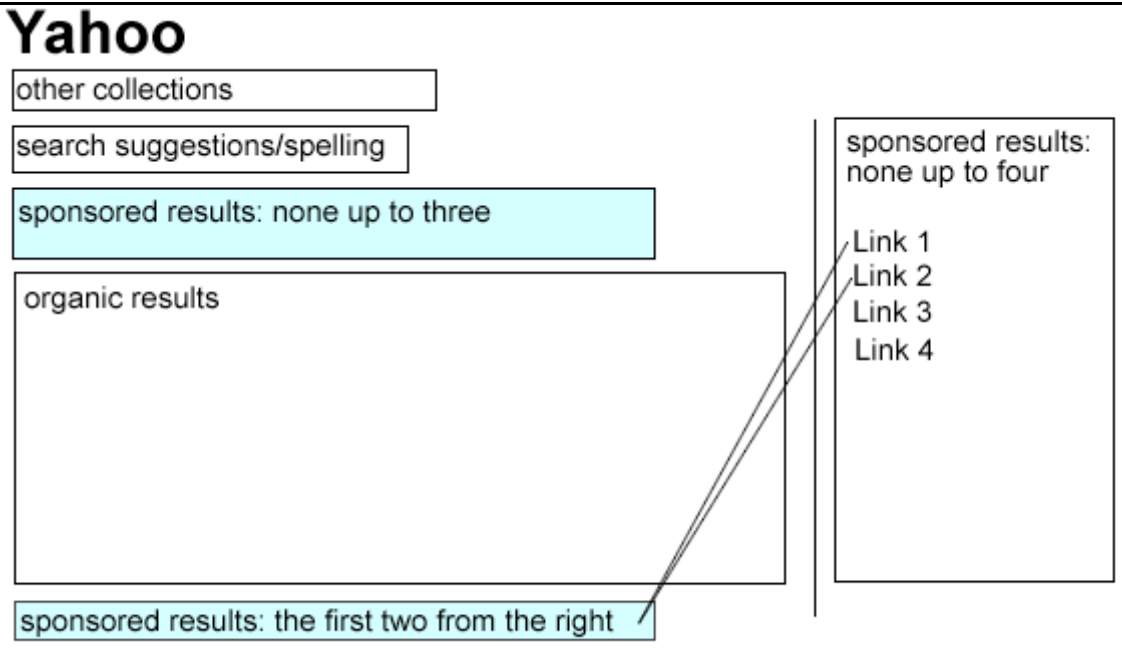

Yahoo has a similar results screen as Google. The only difference is the order of search suggestions and the first sponsored results. First, Yahoo gives some search suggestions, which are followed by sponsored results. The first two sponsored results from the right-hand side also appear below the organic results. On the right-hand side is a maximum of eight sponsored links.

Table 2 summarises the elements used to compose SERPs. From this brief discussion, we can see that SERPs are much more than just organic results and ads. We conclude that, for evaluation purposes, all elements presented should be considered to achieve a realistic view of what the engines actually present to the user.

Table 2: Elements on SERPs

| Name | Description | Position |
|---|---|---|
| Organic | Results from Web crawl. "Objective hits" not influenced by direct payments. | Central on results page. |
| Sponsored | Paid results, separated from the organic results list. | Above or below organic results, on the right-hand side of the results list. |
| Shortcuts | Emphasised result pointing to results | Above organic results, within organic |
| | from a special collection. | results list. |
| Primary search result | Extended result that points to different collections. It comes with an image and further information. | Above organic results, often within organic results. |
| Prefetch | Result from a preferred source, emphasised in the results set. | Above or within organic results. |
| Snippet | Regular organic result with result description extended by additional navigational links. | Within organic results list (usually first position only). |
| Child | Second result from the same server with a link to further results from same server. | Within organic results list; indented. |

**Literature review**

In this section, we give an overview of studies on search engine user behaviour, the result presentation in search engines, and retrieval effectiveness studies. We will show how these three areas influence the results presentation in search engines.

*Search engine user behaviour*

Most searching persons evaluate the results listings very quickly before clicking on one or two recommended Web pages [9, 33]. In a large online survey, nearly a third of 6000 participants complaint about paid listings and the nebulosity of the ranking of results [32]. Paid placements are often not clearly separated from the organic lists. They highlight those links with very light background colours (e.g., Google) or give only hints written in very small and slightly coloured letters (e.g., Altavista). Especially LCD screens do not show a clear contrast and dependent on the ankle of the screen, sometimes no colouring is visible at all. That is why users often cannot differentiate between those two, or have the feeling that the link they clicked on could be a paid listing. Additionally, it is important to present only a few results (a maximum of 10 to 15 links in the visible area) since search engine users are not willing to scroll down [9]. More than half of search engine users only look at the first results page. Remarkably, about half the users submit only one query per session. That indicates the joint occurrence of need for information followed by a search query.

Search queries are very short and do not show any variations over a longitudinal period. Nearly half of the search queries contain only one term. On average, a query contains between 1.6 and 3.3 terms, depending on the query language [10, 14]. In most studies, the most popular search terms or queries are referred to as well. The most popular search queries and terms involve sex or related topics and, in most cases, consist of only one term. Regarding search terms that occur nearly every day [31], one finds many operators used inappropriately and fillers such as “in” or “for”. This shows how intuitively online searching persons formulate their queries. An overview of different studies is given by Höchstötter and Koch [10].

Machill, Neuberger, Schweiger, and Wirth’s study [27, 28] consists of two parts, namely a computer-assisted telephone survey with 1000 interviewed persons and a laboratory study with 150 subjects. They show that 14 percent of search engine users definitely use advanced search features. Only 33 percent of respondents know that it is possible to personalise search engine interfaces. The title and the description of recommended Web pages are very important for users to evaluate the results lists. The conclusion was that search engine users want their searches to be rewarded with success, a quick presentation of results, and clearly designed results screens. It is also remarkable that only a few users use all possibilities, which is why search engines have to serve “sophisticated” results without the users’ help, but with the knowledge search queries in the past.

A shortcoming of most logfile-based studies is that it is only reported how many results pages are viewed by users, but not how many results were selected or in which sequence they click on results. It is common that users in most cases view only the first results page. However, only additional (qualitative) research can show us the results (positions) that are really clicked.[1]

Eye-tracking studies (e.g., [4, 26, 30] show interesting heatmaps, which visualise the examination of search results pages in total and over time. Hotchkiss [8] discussed the Google triangle, which clearly show a user’s focus on the upper left corner. But nowadays, with all these additional pictures and schematic graphics, the user becomes more confused and does not know where to look first. However, the graphical information is always processed before the textual information.

[1] Surely, search engine providers themselves can (and sometimes do) add forwarding links that first point to their own servers to measure clicks (as is also done with sponsored links) and, after that, transfer the user to the selected URL.

Unfortunately, there is little research on the use of other elements presented on SERPs. To our knowledge, there is only some research on the acceptance of query refinement tools on search engines, namely one study discussing the long-gone AltaVista Prisma feature [2] and one dealing with Dogpile.com [15]. Both studies find that users apply the refinement tools only moderately.

Regarding sponsored results, there are some studies. In one study [13], 56 participants evaluated the results from 330 e-commerce result queries in the results screen of a major search engine. A fictitious search engine was created to measure the appeal of sponsored links, the researchers flipped sponsored links and organic results on half of the pages. This meant that study participants who thought they were evaluating organic links sometimes were viewing sponsored links and vice versa. On more than 80 percent of the searches, study participants went first to the results identified as organic. Sponsored links were viewed first for only six percent of the time. Generally, it is more likely that users ignore sponsored results than clicking on them.

In all studies many different aspects have been discussed, but there is no publication at all regarding the results pages as a whole complex with all possibilities of results presentation. To understand what users really see, it is important to draw the complete picture and not only to pick some aspects out of the results pages.

*Results presentation*

While search engines usually return thousands of results, users are not willing to view more than a few [14, 16, 26]. This is the reason search engines de facto have only the first page to present relevant results. This first results page must also be used for presenting results from additional collections, as users usually do not follow links to these additional collections (the so-called "tabs"). Search engine expert Danny Sullivan even coined the term "tab blindness" for this phenomenon.

Results beyond the organic surely need space on the results page. With the addition of more and more results from special collections (and, in some cases, the addition of more and more ads above the organic results), we see a general change in results presentation [20]. Organic results become less important as additional results take their space. Nicholson et al. [29] calculate the "editorial precision" (EPrec) of SERPs by dividing the total screen space by the space used for editorial (i.e., organic) results. This shows that the ratio of space used for organic results differs from search engine to search engine. Apart from the methods used, the results from the study are unfortunately outdated, as the authors use a screen size of 800x600.

*Retrieval effectiveness studies*

Many studies have dealt with the retrieval effectiveness of Web search engines (e.g., [6, 7, 17, 21, 35]. However, the problem with all these studies is that they first only take into account organic results, which means that sponsored results and the results from additional collections are omitted. Some research has been conducted on the effectiveness of sponsored versus organic results [11-13] but, unfortunately, only for queries with a commercial purpose. However, it would be interesting to see whether commercial results are shown for "non-commercial" queries and if so, how the users would judge them.

Second, search engine evaluations generally follow a "TREC-style" approach. This means that the user model is that of a "dedicated searcher" who follows links on the results list in the order presented. In addition, all results are clicked. However, results descriptions heavily influence the decision in favour of or against a result [22]. Retrieval effectiveness tests with a wider focus (regarding results types and the probability that a certain results would be clicked) could help to better understand all the links presented on the first results screen.

**Research objectives**

The above discussion showed that studies conducted so far concentrate on the organic results (and sometimes on the sponsored results). Nicholson et al.'s [29] study is exceptional in that it asks how much space is used for which type of result. We continue working on that question by taking into account all the elements of SERPs described above.

*Research questions*

For the further discussion, we defined the following research questions (RQ):

RQ1: How many sponsored links are on the results screen?

RQ2: Are there popular hosts, domains, and content types preferred by a certain search engine?

RQ3: Is there a difference between search engines regarding the results types presented?

RQ4: How many specially displayed results are on the first results page?

RQ5: To what extent are shortcuts used on the results pages?

RQ6: What is the difference between search engines regarding the questions above?

## Research design

### *Data collection*

For our comparison, we selected five search engines based on the restriction that they should provide their own index and that they are of significance (as expressed in a considerable market share). We chose Google, Yahoo, MSN/Live.com, and Ask.com. This selection corresponds with our other empirical studies on certain aspects of search engine quality [18, 19, 21, 22, 25].

We got 500 queries from the 100k top queries and 500 from the last 100k queries of long tail queries from Ask.com from July 2008. As we had access to the aggregated list of search queries with their occurrences in the query log, it was possible to get a sample containing both popular and rare queries. We sorted the list of all queries by frequency and alphabetically. Then, we selected every second hundredth query. With that, we make sure that we have a representative selection over the popular and over the very rare search queries in the long tail.

In the next step, we wrote a script to automatically download the search engines' results pages. We then developed programs for every search engine mentioned above to analyse the HTML code. We extracted patterns in the HTML code, which gave us the ability to categorise every single result in those pages into organic, paid advertisements, snippets, etc. Some problems occurred with Yahoo, which blocked us after approximately 30 search queries. We then changed the program, so there were only 25 search requests and, afterwards, a 2-hour timeout, until the next request was started. Search engines do not allow people to send machine requests. They state in their policy that they could block one forever or only for a while, whenever they detect such an automatic process. This is understandable, since these major search engines have to protect themselves, especially the usage of sponsored links. They have to interdict automatic requests by robots and scripts; otherwise, it would be possible for one to write a robot that clicks on all sponsored links. We used the data for this research project only, and we did not generate any automatic clicks on any links in the results screen. For Google, we needed to use a proxy in the United States. Google always tracks the IP of its users and always brings up sponsored links of the country identified by the IP. Since we are located in Germany, only German sponsored links would have come up if we did not use the proxy. That is why we used a US proxy and the US web search interface for all search engines. For a different country search one will see different number of results. English search queries wouldn't bring up that many sponsored results in a German web interface and vice versa. For every URL (sponsored or organic) on the results pages, we stored the following information:

- *Position of Organic Results*: position of the URL within the set of organic results
- *Absolute Position Within Complete Set*: the position considering the whole set of elements (including additional results, such as sponsored results or shortcuts shown above the organic results), while *results position* only considers results in the organic results set
- *Absolute Position of Adwords*: position of sponsored results (above, under, or on the right side)
- *URL:* all URLs of organic results have been extracted and most URLs[2] from sponsored results, too.
  *Type of Result*: as stated in Table 2, e.g., organic, sponsored, and shortcut (Whenever we found a shortcut, we also extracted the category of this shortcut, e.g., books, flight, and dictionary
- *Ask 3D*: Ask.com was slightly different, since we also have the information on which terms triggered an Ask 3D result

Therefore, we are able to reconstruct all elements shown on the results list, including their position. We can model the elements shown for the different screen and browser window sizes, respectively.

We extracted all data and shortcuts found in our examination. We will compare the results regarding the top queries and the queries from the heavy tail.

[2] Google's Adwords had been masked by the proxy, which is why we did not extract the URLs of sponsored links for Google, but only the position and number.

**Results**

Since every search engine in our experiment has different modules to be presented as discussed above, we will present the analysis of results screens for every search engine separately. We will also analyse the most popular hosts, domains, top-level domains (TLDs), and file types. Those will be compared directly. The overview of sponsored and organic results in popular and rare queries will also be discussed to give an overview.

The Google result set contains 499 popular search queries and 498 queries from the heavy tail. Those search queries generated 12,522 results in total. The popular search queries come up with 6,731 results, and the rare ones, with 5,791. Only 16 rare search queries generated no results at all.

Yahoo produced 463 results sets for the popular queries and 492 for the rare queries, respectively. In total, 9,436 results from Yahoo were processed, where 5,232 were generated from popular queries, and 4,204, from heavy-tail queries. Sixty-four of these produced no results at all, while only 2 popular queries had no results.

We obtained 11,752 results from the MSN search engine. All 500 popular search queries produced results (6,685). Only 457 of the rare results are valid; all other search queries had been blocked or did not go through for some other reason. Of those rare queries, only 12 did not generate any results. We got 5,065 results from rare queries.

From Ask.com, we obtained a total of 9,127 URLs. All popular queries were processed, while 2 rare queries could not be processed. Popular queries produced a total of 5,224 results, and rare queries, a total of 3,903 results. Five popular queries led to an empty results set, while for the rare queries, the amount was 79. Table 3 gives an overview of the results sets of all search engines under investigation. It also clearly shows how many sponsored links are on the first results screen.

Table 3: Overview of the Results Sets for the Search Engines Investigated

| | Google | Yahoo | MSN/Live | Ask |
|---|---|---|---|---|
| Valid popular queries | 499 | 463 | 500 | 500 |
| Valid rare queries | 498 | 492 | 457 | 498 |
| URLS in results screens | 12,522 | 9,436 | 11,700 | 9,127 |
| URLs (from popular) | 6,731 | 5,232 | 6,685 | 5,224 |
| URLs (from rare) | 5,791 | 4,204 | 5,065 | 3,903 |
| Organic URLs | 9,641 | 8,454 | 9,177 | 8,183 |
| Organic URLs (from popular) | 5,041 | 4,543 | 4,996 | 4,661 |
| Organic URLs (from rare) | 4,600 | 3,911 | 4,181 | 3,522 |
| Sponsored Links | 2,881 | 982 | 2,573 | 944 |
| Sponsored Links (from popular) | 1,690 | 689 | 1,689 | 563 |
| Sponsored Links (from rare) | 1,191 | 293 | 884 | 381 |
| No results (from popular) | 0 | 2 | 0 | 5 |
| No results (from rare) | 16 | 64 | 12 | 79 |

One can clearly see that there are differences in the amount of sponsored results regarding rare and popular queries. Popular queries come clearly with more sponsored Links.

*Characterisation of the Organic Results Sets*

In this section, we will characterise the organic results by their content, TLD, hosts, and preferred file types. The dataset contains 9,641 organic results from Google, 8,454 from Yahoo, 9,177 from MSN, and 8,183 from Ask.com.

Table 4: TLD Distribution Within the Results Sets

| Top | Google | | | Yahoo | | MSN/Live | | Ask | |
|---|---|---|---|---|---|---|---|---|---|
| 1 | .com | 6614 | 68.6% | 5519 | 65.3% | 5255 | 57.3% | 4450 | 54.4% |
| 2 | .co.uk | 1141 | 11.8% | 830 | 9.8% | 1350 | 14.7% | 1630 | 19.9% |
| 3 | .org | 1029 | 10.7% | 1014 | 12.0% | 1070 | 11.7% | 687 | 8.4% |
| 4 | .net | 300 | 3.1% | 382 | 4.5% | 304 | 3.3% | 271 | 3.3% |
| 5 | .gov | 130 | 1.3% | 111 | 1.3% | 89 | 1.0% | 100 | 1.2% |
| 6 | .edu | 141 | 1.5% | 170 | 2.0% | 231 | 2.5% | 166 | 2.0% |
| 7 | .gov.uk | 111 | 1.2% | 73 | 0.9% | 140 | 1.5% | 105 | 1.3% |
| 8 | .au | 72 | 0.7% | 56 | 0.7% | 106 | 1.2% | 91 | 1.1% |
| 9 | .info | 59 | 0.3% | 35 | 0.4% | 55 | 0.6% | 17 | 0.2% |
| 10 | .us | 40 | 0.4% | 50 | 0.6% | 21 | 0.2% | 29 | 0.4% |

As anticipated, most results come from the .com TLD. This is not remarkable, as we used the .com interfaces for our study. Search engines heavily bias search results towards the language and the country from which the query comes (Lewandowski, 2008b). While Google and Yahoo rely on .com results the most (with 68.6% and 65.3% of results, respectively), the ratio is lower for MSN and Ask.com (57.3% and 54.4%, respectively). We can see, from the data, that the latter two engines, in contrast, have a higher proportion of results from the .co.uk domain (14.7% for MSN and 19.9% for Ask.com). However, we cannot say whether this leads to better or worse results. Surely, results from the United Kingdom could, in some cases, be relevant for an American user.

Results from the .org TLD account for around 10% of all organic results and vary from 10% for Ask.com to 12% for Yahoo. The .net TLD comes in all search engines before the .gov domain. The .edu results account for around 2% of results, UK government sites (.gov.uk), for around 1% (ranging from 0.9% for Yahoo to 1.5% for MSN). It is remarkable, that there is no real difference between .gov and .gov.uk TLDs in the results. One would expect that, for queries from the United States, .gov results would come up more often than results from UK government sites (For an overview of the most popular TLDs in the search engines under investigation, see Table 4).

Table 5: Host and Domain Distribution of All Organic Results within the Result Sets

| Top | Google | Count | Yahoo | Count | MSN/Live | Count | Ask | Count |
|---|---|---|---|---|---|---|---|---|
| 1 | en.wikipedia.org | 328 | en.wikipedia.org | 446 | en.wikipedia.org | 387 | en.wikipedia.org | 255 |
| 2 | www.youtube.com | 169 | www.shop.com | 102 | dictionary.reference.com | 52 | www.bbc.co.uk | 54 |
| 3 | answers.yahoo.com | 76 | www.NexTag.com | 90 | www.amazon.com | 45 | www.tripadvisor.com | 48 |
| 4 | www.amazon.com | 72 | answers.yahoo.com | 88 | www.google.com | 38 | www.geocities.com | 29 |
| 5 | books.google.com | 67 | www.Calibex.com | 54 | www.imdb.com | 34 | www.ciao.co.uk | 25 |
| 6 | wiki.answers.com | 51 | www.answers.com | 49 | www.youtube.com | 33 | www.kijiji.de | 22 |
| 7 | www.imdb.com | 46 | www.imdb.com | 49 | www.bbc.co.uk | 29 | www.myspace.com | 22 |
| 8 | www.bbc.co.uk | 43 | www.youtube.com | 46 | search.live.com | 21 | www.theanswerbank.co.uk | 21 |
| 9 | news.bbc.co.uk | 33 | www.amazon.com | 41 | groups.google.com | 19 | www.dooyoo.co.uk | 19 |
| 10 | findarticles.com | 32 | www.myspace.com | 37 | www.reviewcentre.com | 19 | www.amazon.com | 18 |
| 11 | www.reviewcentre.com | 28 | tripadvisor.com | 36 | www.tripadvisor.com | 19 | www.nlm.nih.gov | 18 |
| 12 | www.tripadvisor.com | 27 | www.geocities.com | 35 | news.bbc.co.uk | 18 | www.reviewcentre.com | 18 |
| 13 | ezinearticles.com | 26 | search-desc.ebay.com | 32 | profile.myspace.com | 18 | www.amazon.co.uk | 17 |
| 14 | www.answers.com | 25 | www.target.com | 27 | www.answers.com | 17 | members.lycos.co.uk | 16 |
| 15 | Profile.myspace.com | 20 | shopping.yahoo.com | 26 | Us.imdb.com | 16 | www.angelfire.com | 16 |
| | | | | | | | www.creditgate.com | 16 |
| | | | | | | | www.imdb.com | 16 |
| | | | | | | | www.youtube.com | 16 |

Often, users complain about the top results from the popular search engines in that they produce results from the same hosts for the majority of queries. Therefore, a closer look at the distribution of the top hosts might prove useful.

Table **5** gives an overview of the most popular domains in the different search engines. One can see that the most popular host with all search engines is the English-language site of Wikipedia. We will discuss Wikipedia results in detail in the next section.

It is interesting to see how the different search engines deal with some popular hosts. Therefore, we will compare some of them. Results from YouTube are heavily featured within the results lists from Google. This comes as no surprise, as YouTube is a subsidiary of Google. However, there might be good reasons (beyond business) for this. Users might prefer multimedia results and/or a mixed result set. However, none of the other search engines features YouTube results as much as Google does. While in our dataset, Google produces 169 YouTube results, Yahoo only produces 45; MSN, 33; and Ask.com, only 16. Another Google domain, books.google.com, is also prominently featured within Google's results. As with other services provided by Google, one is not allowed to crawl books.google.com. Those results are also dedicated to Google and its partner program with publishers or libraries. It is interesting to see the contrast when looking at the only-Yahoo property found in the top 15 lists. Yahoo Answers (answers.yahoo.com) is prominently featured on Yahoo (85 results) and on Google as well (67 results). The other engines do not place this host prominently.

The commercial results often complained about by users (e.g., [27, 28, 32]) are those from Amazon. These results can be found in the top 15 lists of all search engines under investigation. Google produces by far the most results from this host (72), while Yahoo produces 39; MSN, 45; and Ask.com, 35 (18 from Amazon.com and 17 from Amazon.co.uk).

We close our examination of the popular hosts with another interesting example, namely the highly regarded news site from the British Broadcasting Cooperation (BBC). All search engines except Yahoo produce a considerable amount of BBC results (whether from the news.bbc.co.uk domain or from the www.bbc.co.uk domain): Google gives a total of 76 results from the BBC; MSN, 47; and Ask.com, 54. It is remarkable that there are so many results from the Google domain (google.com) in MSN's results. Those nearly all come from www.google.com/notebook/public/....., which is a service by Google to create a private notebook for collecting interesting sources on the Web while browsing. This function is probably used by search engine optimisers. This could be an explanation as to why those results come up on the MSN results screen. They are not very good, and, most of the time, they are dead links.

The distribution of hosts in search engines also shows heavy tails. In every search engine, the proportion of hosts that only appeared once in the results pages under investigation is larger than 50%. For Google it is 54.7%; for Yahoo, 55.1%; for MSN, 57.9%; and for Ask.com, 58.1%. The percentage of hosts that came up twice varies from 7.6% (Google) to 13.1% (Ask.com). Yahoo and MSN have around 10% of their hosts coming up twice.

By looking at Wikipedia results only (Table 6), we can see that not only is Wikipedia the absolute winner in every search engine but also Google clearly boosts Wikipedia results towards the pole position. It seems that when a Wikipedia result is found for a certain query, it will be shown in one of the first positions. This holds true for the rare queries and for the popular queries to an even larger degree. While Yahoo shows a larger amount of Wikipedia results in total, the distribution over the top 10 results is wider. MSN's strategy seems to be similar to Yahoo's. Only Ask.com is the odd one, with no clear patterns. It seems that there is some boost, but only in the upper and most likely in the visible area. In addition, Ask.com boosts Wikipedia results (to a lesser extent) when Ask 3D is triggered (see Table **14**).

Regarding the visible area, one can clearly see, from the data, the reason users might come to the conclusion that search engines in general and Google in particular "always" show Wikipedia results. When only looking at an assumed visible area showing four organic results, Google shows Wikipedia results for 35% of popular queries and for 18.7% of rare queries (on average). In Table 6, we listed all Wikipedia results from all screens, even if a Wikipedia result is a child of another Wikipedia result. In Google, there are 55 queries with two Wikipedia results, of which 12 are rare and 43 popular. There are never more than two Wikipedia results on the first results page. This is also the case for all the other search engines. However, we did not calculate these numbers for pairs of Wikipedia results in results pages, since the numbers below already give an overview of what the users see at the top. In Google, for 16.2% of all popular results, there is a Wikipedia result in the first position, which is a lot.

Table 6: Distribution of Wikipedia Results in Popular and Rare Queries

| Wikipedia | Google | | Yahoo | | MSN/Live | | Ask.com | |
|---|---|---|---|---|---|---|---|---|
| Position | Popular | Rare | Popular | Rare | Popular | Rare | Popular | Rare |
| 1 | 81 | 53 | 58 | 41 | 62 | 39 | 9 | 9 |
| 2 | 49 | 16 | 59 | 31 | 54 | 31 | 33 | 16 |
| 3 | 25 | 16 | 64 | 21 | 45 | 15 | 20 | 18 |
| 4 | 20 | 8 | 41 | 21 | 30 | 11 | 31 | 12 |
| 5 | 5 | 5 | 26 | 13 | 20 | 12 | 17 | 4 |
| 6 | 9 | 5 | 10 | 9 | 15 | 12 | 17 | 8 |
| 7 | 9 | 4 | 11 | 10 | 8 | 7 | 15 | 12 |
| 8 | 7 | 4 | 6 | 6 | 5 | 6 | 7 | 5 |
| 9 | 4 | 1 | 5 | 7 | 3 | 3 | 10 | 4 |
| 10 | 2 | 4 | 6 | 1 | 2 | 7 | 6 | 2 |
| ***Total*** | ***212*** | ***116*** | ***286*** | ***160*** | ***244*** | ***143*** | ***165*** | ***90*** |

For a better overview we also included a table with the percentages of Wikipedia results (Table 7).

Table 7: Percentages of Wikipedia Results in Popular and Rare Queries

| Wikipedia | Google | | Yahoo | | MSN/Live | | Ask.com | |
|---|---|---|---|---|---|---|---|---|
| Position | Popular | Rare | Popular | Rare | Popular | Rare | Popular | Rare |
| 1 | 38.2% | 45.7% | 20.3% | 25.6% | 25.4% | 27.3% | 5.5% | 10.0% |
| 2 | 23.1% | 13.8% | 20.6% | 19.4% | 22.1% | 21.7% | 20.0% | 17.8% |
| 3 | 11.8% | 13.8% | 22.4% | 13.1% | 18.4% | 10.5% | 12.1% | 20.0% |
| 4 | 9.4% | 6.9% | 14.3% | 13.1% | 12.3% | 7.7% | 18.8% | 13.3% |
| 5 | 2.4% | 4.3% | 9.1% | 8.1% | 8.2% | 8.4% | 10.3% | 4.4% |
| 6 | 4.2% | 4.3% | 3.5% | 5.6% | 6.1% | 8.4% | 10.3% | 8.9% |
| 7 | 4.2% | 3.4% | 3.8% | 6.3% | 3.3% | 4.9% | 9.1% | 13.3% |
| 8 | 3.3% | 3.4% | 2.1% | 3.8% | 2.0% | 4.2% | 4.2% | 5.6% |
| 9 | 1.9% | 0.9% | 1.7% | 4.4% | 1.2% | 2.1% | 6.1% | 4.4% |
| 10 | 0.9% | 3.4% | 2.1% | 0.6% | 0.8% | 4.9% | 3.6% | 2.2% |
| Total | 100.0% | 100.0% | 100.0% | 100.0% | 100.0% | 100.0% | 100.0% | 100.0% |

The most common file types on results screens are HTML and HTM pages ( Table **8**). We listed both separately since there is an important distinction: The HTM ending could give a hint to UNIX systems. Those cannot use the HTML file extension. No search engines like dynamic file types, which is understandable. It may be that the search engine's crawler collects a huge database that is dynamically generated.

Ask.com obviously has a problem with indexing Microsoft office documents. In addition, PDFs are more seldom to be found than in the competitors' SERPs. All other engines do have those file types in their SERPs but not that often. HTML/HTM is the file type of the Web and definitively easier to crawl, tag, and index than other file types. This could be an explanation as to why search engines mostly retrieve HTML/HTM results from their indexes. The signals (titles, descriptions, etc.) of those documents are probably better and well structured.

Table 8: Common Filetypes in Organic Results

| Type | Google | Yahoo | MSN/Live | Ask.com |
|---|---|---|---|---|
| .html | 1,838 | 1,423 | 1,366 | 1,982 |
| .htm | 885 | 787 | 899 | 1,149 |
| .php | 176 | 182 | 221 | 153 |
| .pdf | 158 | 152 | 111 | 50 |
| .aspx | 141 | 130 | 152 | 97 |
| .asp | 140 | 140 | 199 | 126 |
| .shtml | 111 | 84 | 87 | 106 |
| .cfm | 30 | 32 | 42 | 18 |
| .doc | 29 | 5 | 11 | 0 |
| .ppt | 4 | 2 | 2 | 0 |

*Use of Advertisements for Popular and Rare Queries*

From the Google results, we extracted 12,522 results in total, where 9,641 results are organic, and the rest of the 2,881 results are sponsored. This means that 77% of all URLs on the results screens are organic, and 23% are sponsored. We got 9,436 results from Yahoo, with 9,436 organic results and 982 sponsored results, which is much less than in Google, since 89.6% of all results are organic. From MSN/Live, 11,700 URLs have been extracted, and the portion of organic results is 78.4%. Ask.com shows 944 sponsored results in 9,127 results retrieved total. This is a portion of 89.7% of organic results. These results clearly show that the market leader Google is also the leader in bringing up sponsored results. As we assumed before, popular queries have more sponsored links on their results screens than rare queries (see Table 9 and Table **10**). This is absolutely understandable, since an advertiser will try to get traffic over the search queries. The probability of getting more traffic increases with the frequency of a search term. MSN and Google have more advertisements than the other search engines.

In Table 9, the use of sponsored links for popular queries can be seen. The table shows in how many cases advertisements are placed at a certain position on the results pages of the different engines. Not only can we see in which areas of the screen the ads are shown (top, right, and below; see the schematic representation in Table 1) but also the different strategies of the search engines. When looking only at the ads presented above the organic results lists, Yahoo and MSN have a higher percentage of queries where ads are shown at this position (38%), while ads show up at this position at a lower ratio on Google and Ask.com (approximately 27%). However, the strategies of Google and Ask.com are different: Google's strategy is to rely on ads on the right-hand side of the results screen. It is most likely that a search query has at least one sponsored link on its SERP. On Google, 297 of all popular queries have at least one sponsored link on the page, and 230 of the SERPs are generated by rare queries. So, over half of popular queries have advertisements on the SERPs. However, the lower amount of ads presented above the organic results could lead users to think that this search engine shows fewer ads in general.

In contrast to the other search engines, Ask.com has no advertisements on the right-hand side of the organic results at all. However, this search engine shows more ads above the organic results, which could lead to a visible area without any organic results at all. Ask.com only has sponsored links on the top and bottom. This could lead to the case where a user scrolls down the results page and clicks on a sponsored link accidentally, but this case is not very likely. This could be the reason the sponsored links on top are the same as those on the bottom and that there are more sponsored links at the very end. The distribution of the upper and lower ads is slightly different, but is more likely to have full five sponsored links on the bottom than on the top.

Table 9: Use of Advertisements for Popular Queries

| | | Google | | Yahoo | | MSN/Live | | Ask | |
|---|---|---|---|---|---|---|---|---|---|
| Top | all | 137 | 27.4% | 180 | 38.9% | 193 | 38.6% | 134 | 26.8% |
| | 1 | 51 | 10.2% | 120 | 25.9% | 70 | 14.0% | 63 | 12.6% |
| | 2 | 26 | 5.2% | 50 | 10.8% | 39 | 7.8% | 33 | 6.6% |
| | 3 | 60 | 12.0% | 10 | 2.2% | 84 | 16.8% | 13 | 2.6% |
| | 4 | None | | None | | None | | 14 | 2.8% |
| | 5 | | | | | | | 11 | 2.2% |
| Right | all | 287 | 57.5% | 117 | 25.3% | 329 | 65.8% | None | |
| | 1 | 62 | 12.4% | 58 | 12.5% | 107 | 21.4% | | |
| | 2 | 32 | 6.4% | 25 | 5.4% | 52 | 10.4% | | |
| | 3 | 24 | 4.8% | 13 | 2.8% | 36 | 7.2% | | |
| | 4 | 17 | 3.4% | 7 | 1.5% | 16 | 3.2% | | |
| | 5 | 13 | 2.6% | 5 | 1.1% | 118 | 23.6% | | |
| | 6 | 13 | 2.6% | 2 | 0.4% | None | | | |
| | 7 | 10 | 2.0% | 5 | 1.1% | | | | |
| | 8 | 116 | 23.2% | 2 | 0.4% | | | | |
| Below | 1 | None | | 58 | 12.5% | 70 | 14.0% | 63 | 12.6% |

| | n | Yahoo | | MSN/Live | | Ask | |
|---|---|---|---|---|---|---|---|
| | 2 | 59 | 5.4% | 123 | 24.6% | 33 | 6.6% |
| | 3 | None | | None | | 13 | 2.6% |
| | 4 | | | | | 9 | 1.8% |
| | 5 | | | | | 16 | 3.2% |

One can see that the distribution of the number of sponsored links on top has no clear trend. Ask.com has a clear trend, since there are no sponsored results on the right, and sponsored links come from Google. It is likely that a search query generates a results page with one sponsored link on the top and also very likely that a search query triggers two sponsored links on top. This is not very clear, but obviously search queries trigger mostly one or three top sponsored links.

A similar pattern corresponds to the sponsored links on the right side. There is a huge increase of queries with eight sponsored links. An explanation for both could be that a query is used to get as many sponsored results as possible. Everything in between depends on the interest of advertisers in the keywords. Another reason is the distribution of booked terms. There will be a huge portion of popular terms to book. But the more advertisers are booking those terms, the more sponsored links will come up. This is probably the best explanation for the increase in queries with eight sponsored links on the right or three sponsored links on the top. These numbers give an impression of what users will see in the visible area of their screen.

Table 10: Use of Advertisements for Rare Queries

| | | Google | | Yahoo | | MSN/Live | | Ask | |
|---|---|---|---|---|---|---|---|---|---|
| Top | all | 92 | 18.4% | 99 | 20.1% | 109 | 23.9% | 86 | 17.3% |
| | 1 | 43 | 8.6% | 77 | 15.7% | 46 | 10.1% | 34 | 6.8% |
| | 2 | 15 | 3.0% | 19 | 3.9% | 25 | 5.5% | 21 | 4.2% |
| | 3 | 34 | 6.8% | 3 | 0.6% | 38 | 8.3% | 16 | 3.2% |
| | 4 | None | | None | | None | | 10 | 2.0% |
| | 5 | | | | | | | 5 | 1.0% |
| Right | all | 227 | 45.6% | 53 | 10.8% | 188 | 41.1% | None | |
| | 1 | 56 | 11.2% | 31 | 6.3% | 61 | 13.3% | | |
| | 2 | 27 | 5.4% | 10 | 2.0% | 45 | 9.8% | | |
| | 3 | 33 | 6.6% | 5 | 1.0% | 24 | 5.3% | | |
| | 4 | 10 | 2.0% | 5 | 1.0% | 11 | 2.4% | | |
| | 5 | 10 | 2.0% | 2 | 0.4% | 47 | 10.3% | | |
| | 6 | 3 | 0.6% | 0 | 0.0% | None | | | |
| | 7 | 5 | 1.0% | 0 | 0.0% | | | | |
| | 8 | 83 | 16.7% | 0 | 0.0% | | | | |
| Below | 1 | None | | 58 | 12.5% | 70 | 14.0% | 30 | 6.0% |
| | 2 | | | 59 | 5.4% | 123 | 24.6% | 21 | 4.2% |
| | 3 | | | None | | None | | 8 | 1.6% |
| | 4 | | | | | | | 4 | 0.8% |
| | 5 | | | | | | | 16 | 3.2% |

In an average Google results screen for popular queries, a user sees 0.57 sponsored results on top and 2.8 results on the right side. These numbers are smaller for rare queries. There, only 0.35 sponsored results will be on top, and 2.0 will be on the right side. For popular queries, Yahoo shows only 0.53 top sponsored results and 0.57 on the right side. There are much fewer right sponsored results since Yahoo does not increase the portion of queries that trigger the full amount of sponsored links. The average for rare queries is 0.25 on top and 0.2 on the right side. MSN has 0.8 on top and 1.9 on the right side for popular queries. Rare queries' results pages will be covered by 0.45 on top and 1.1 on the right. Ask.com shows only 0.56 top sponsored results for popular queries and 0.38 for rare queries. The average number only increases slightly when queries with no results at all are omitted. For Ask.com, users would see 0.45 sponsored results on top for rare queries without them.

*Use of Shortcuts for Popular Queries*

In this section, we take a closer look at the use of shortcuts by the different search engines. As the search engines use different shortcuts (and in a different way), we discuss each engine separately.

Table 11: Use of Shortcuts for Popular and Rare Queries in Google

| | | Popular | | Rare | |
|---|---|---|---|---|---|
| | | Count | Position | Count | Position |
| | Prefetch | 170 | 1 | 69 | 1 |
| | Snippet | 290 | var | 196 | var |
| | Images | 9 | 1 | 5 | 1 |
| Shortcuts | ***Total*** | ***183*** | **-** | ***60*** | **-** |
| | Local results | 2 | 1 | 4 | 1 |
| | Video | 108 | var | 127 | var |
| | Blog search | 6 | 11+ | 4 | 11+ |
| | Books | 22 | 10+ | 9 | 10+ |
| | Calculator | 7 | 1 | 1 | 1 |
| | Dictionary | 1 | 1 | 3 | 1 |
| | News | 10 | 3/10 | 4 | 3/10 |
| | Scholar | 3 | 1 | 2 | 1 |
| | Shopping | 17 | var | 37 | var |
| | Weather | 5 | 1 | 1 | 1 |
| | NASDAQ | 2 | 1 | - | - |

In Google, of all 499 popular search queries, 373 (74.7%) had at least one of the results listed above (see Table **11**). This means that every popular search query that triggers one of those results has 1.7 smart results, on average. Of all 498 rare search queries, 297 (59.6%) got at least one of those special results; this is 1.5 on average. If one takes all search queries into account, popular search queries will have 1.3 smart results per query; and rare queries, 0.9. It is remarkable that news results are always in the third or tenth position in the organic results list. Snippets and prefetches are not found that often in the results set for rare queries, but there are more than twice that of shopping links. The results show that Google learned from what users searched for every day in the past. Google is also more flexible with the positions of shortcuts. A shortcut in Google results is not always on the pole position.

Table 12: Use of Shortcuts for Popular and Rare queries in Yahoo

| | | Popular | | Rare | |
|---|---|---|---|---|---|
| | | Count | Position | Count | Position |
| | Prefetch | 45 | var | 16 | var |
| | Snippet | 25 | 1 | 10 | 1 |
| Shortcuts | ***Total*** | ***49*** | var | ***39*** | |
| | Images | 11 | 1 | 8 | 1 |
| | Video | 24 | 1/2 | 28 | 1/2 |
| | News | 12 | 1 | 2 | 1 |
| | Local results | 0 | | 1 | 1 |
| | Travel | 2 | 1 | 0 | - |

Yahoo has nearly the same distribution of shortcuts for popular and rare queries (see Table 12). However, Yahoo is very "under-sophisticated" with respect to its shortcuts in all SERPS. There are only a few categories. But there are some prefetches and snippets. However, Yahoo does not show as many results from other collections and shortcuts as the other search engines do.

Table 13: Use of Shortcuts for Popular and Rare Queries in MSN

| | | Popular | | Rare | |
|---|---|---|---|---|---|
| | | Count | Position | Count | Position |
| | Snippet | 37 | 1/2 | 13 | 1/2 |
| Shortcuts | ***Total*** | ***44*** | | ***17*** | |
| | Images | 12 | 1 | 6 | 1 |
| | Videos | 2 | 1 | 1 | 1 |
| | News | 12 | 1 | 2 | 1 |
| | Local results | 3 | 1 | 2 | 1 |
| | Encarta | 8 | 1 | 2 | 1 |
| | Xrank™ | 3 | 1 | 0 | 1 |
| | Products | 4 | 1 | 4 | 1 |
| Top answer | ***Total*** | ***1*** | | ***7*** | |
| | Encyclopedia | 1 | 1 | 6 | 1 |
| | Stock exchange | - | - | 1 | 1 |

MSN/Live is also much underrepresented with respect to its shortcuts in all SERPS (see Table 13). There are many more categories, but they do not come up that often. There is also a difference in popular and rare queries, but even the popular ones do not trigger as many shortcuts as they do on Google or Ask.com.

Table 14: Use of Shortcuts for Popular and Rare Queries on Ask.com

| | | Popular | | Rare | |
|---|---|---|---|---|---|
| | | Count | Position | Count | Position |
| Ask 3D | | 157 | - | 78 | - |
| Primary SR | | 26 | 1 | 4 | - |
| News injected | | 22 | 1-4 | 13 | - |
| Shortcuts | ***Total*** | ***35*** | | ***14*** | |
| | Currency converter | 3 | | 2 | |
| | Unit converter | 9 | | 0 | |
| | Wikipedia | 14 | | 3 | |
| | Local results | 0 | | 3 | |
| | Real estate | 0 | | 1 | |
| | Dictionary | 1 | | 5 | |
| | News | 3 | | 1 | |
| | Gifts | 1 | | 0 | |
| | RSS | 1 | | 0 | |
| | Weather | 2 | | 0 | |
| | Smart answer | 1 | | 0 | |

Ask.com always has its shortcut links on the top of the results list (Table 14). The only shortcuts that are injected into the main organic result list are news items. But those are different from shortcut news items on the top. This is interesting, since no other search engine is doing the same. It may be that there are two news results, one from the shortcut and one from the injection. Ask.com has the largest portfolio of shortcuts that can come up on their results screens. As with the other competitors, popular search queries have more shortcuts in their SERPs. One has to look differently at those results, since there is also ASK 3D and the primary search result. Whenever there are Ask 3D results, different results from other collections show up automatically. But this can also cause redundancies, since, e.g., several news results could show up.

However, Google has the largest amount of “sophisticated” results in its results screen. Google knows the preferences of its users very well and tries to help by bringing up those extended results. Google was also a step ahead by implementing some of those results first. Google was able to show local results earlier than all other competitors, since it already had its GoogleMaps service running. Google has a large number of additional services and can use them to infiltrate a variety of results on its results pages.

**Discussion**

Our results show that it is all about the first results screen in search engines – and they know it. Every search engine tries to improve the presentation of results on the first results page by injecting different kinds of results. They also try to promote their own other collections, such as image searches and news searches. It is also true that users do not use those collections when not explicitly pointed to them, but they will become lazy, since they already have other results on the first page. And users are lazy as older studies show regarding the clicking behaviour on the first result screen and the usage of advanced search features and indexes [8, 10]. The first page is not about pure Web results anymore; it is about multimedia results. It is also remarkable that every search engine is predicting the need behind their users' queries. From our data, one can clearly see that search engines favour some hosts in their (top) results. This is perfectly natural, as some hosts are good resources and should therefore come up in the results lists. However, it is at least problematic when a certain search engine clearly favours results from one of its own subsidiaries. We found that Google shows results from its YouTube service to a disproportional degree when compared to the other engines.

The preference for some hosts (as Wikipedia) could lead to results sets in the visible area, where most of the space available, filled with boosted results instead of "real" organic listings. Together with shortcuts (that, at least in some cases, lead to partner sites of the search engine itself) and top results often coming from highly optimised Web sites (taken care of by search engine optimisers), search engines are far from neutral results lists. One could argue that search engines are not bound to produce neutral results [36]; one point is that users could easily switch search engines). However, users should be aware of the composition of SERPs and, then, decide whether to trust them.

Our study gives good material for the badly needed discussion on the influence of search engines on the acquisition of information (and knowledge). Taking into account both the market dominance of only a few search engines and the biased results pages, we can see the real power of certain search engines in determining what users see.

Our study also shows interesting implications for search engine optimization, since the first listing are already taken by good preselected hosts and Wikipedia result, it is, in general, not possible always to take the first listings. Especially for popular terms this undertaken is quite difficult.

Our research also has implications for search engine retrieval effectiveness tests. While these tests can model the results positions and the position where the user looks no further at results (precision@n), we can now model the number of results seen directly by the user (visible area) and consider all elements on the results pages. Retrieval effectiveness tests should consider all elements presented, as user need could well be satisfied with results from additional collections or shortcuts. Therefore, a restriction to the organic results list introduces a bias in such tests.

**Conclusion**

With these results one gets an overview of the distribution of sponsored results in search engine results pages. Google is the winner in showing sponsored results, with a quarter paid listings of all URLs in their first results pages. Since Ask uses Google sponsored links, their results are similar. Sponsored results do not come up that often with rare queries (RQ1).

The Preferences for some TLDs are similar for all search engines. Of course the .com TLD is on the first place. There are definitively some popular hosts being reused on results pages and there are differences in preferences of the top four search engines. Especially Wikipedia plays an important role. Here, Yahoo and MSN place the most Wikipedia results on their results pages. Google boost Wikipedia result mostly on first position but shows less Wikipedia links in total. This is interesting, since Wikipedia is a well-known web site and known by every search engine, so, this shows differences in their ranking algorithms. A clear difference between popular and rare queries is quite obvious (RQ2).

The distribution of file types is similar for all those search engines. HTLML/HTM is the most popular type. This shows that all search engines are still web search engines based on HTML-content in their indexes. Other file types are not that present which is clear, since HTML-code gives the most possibilities for information retrieval, much more than images or videos (RQ4).

Google is on the first place in special results representation in their results pages. No other search engine has that many different types that often on their pages. But the frequency differs for rare and popular queries (RQ5).

Regarding those results and the possibilities given by search engines, one can see that there is a preference in search queries and that the probability of search queries being used and the number of special results varies with that preference. Even though tendencies are the same for all search engines especially the usage of sponsored and special results shows differences. But popular search queries are definitively preferred (RQ6).

**Further Research**

Our research is limited in that we only used 1000 queries per search engine, 500 popular and 500 rare. This means that we can only give evidence for such queries. In addition, the portion of rare queries with no results differs from engine to engine. However, the distribution of search queries is generally highly skewed [10, 32]. Therefore, further research should also focus on the middle part of the distribution. This part is located in the heavy-tail area as well, but those search queries have frequencies from two and above, per month. Interestingly, there was a difference between "real" heavy-tail search queries, which only have a frequency of one, and those queries that appear very often and are under the top 100k. We have to expand this investigation by search query categories, such as commercial or multimedia queries. Also the exploration of navigational, transactional and informational queries will be interesting. Another perspective on this topic is to compare those results regarding different length of search queries.

As with all search engine research using queries (and interfaces) in one language, results can only give tendencies for other languages. This should be kept in mind when extrapolating results to other languages. In further research, we will apply our methods to other languages. Here, there will be a change in the arrangement of results pages, too. To give an example, Ask.com does not bring up Ask 3D for any European country.

To further extend the usefulness of this research, it is necessary to continue monitoring the SERPs for changes in presentation and bias towards certain hosts, whether through the top positions in the organic results lists or through shortcuts. From our data, we will also calculate overlaps between the different search engines. There are some overlap studies (e.g., [34]), but, with the results presentation shown in this paper, more sophisticated methods than simply checking the top 10 organic URLs must be applied. It will also be challenging if there are differences by taking the length of queries into consideration. There are probably differences in the appearance of sponsored links, shortcuts, and snippets as well.

Our findings show the results presentation on normal computer screens. However, results presentation will change, as more and more queries will be entered on mobile devices. Mobile phone screens are a challenge for search engines: There is only space for the organic results, not to mention space for the presentation of advertisements. We will have to wait to see what solution will become the standard. It will be interesting to see how the "conflict" between organic and sponsored results will be solved on such a small screen.

One often hears that searching is all about relevance. This may be true when only the users' needs are considered. However, search engines make money from advertising, and other forms of revenues are, at least today, of no importance. Therefore, search engines must find a good compromise in the presentation of both types of results. While contextual ads are not necessarily irrelevant [11], organic results (and neutral, i.e., non-paid shortcuts) are clearly preferable to the user.

With these generally results it will be interesting to investigate further what the user see most often and which combinations does he see most likely. It will be also a task to investigate different categories and length of queries and their results screens. We assume that longer queries will bring up less paid listings and commercial search queries will produce more sophisticated results. It will be interesting to see the most popular scenario regarding all results in SERPs, screen sizes and visible area to reflect what users see.